\documentclass{article}

\usepackage{upgreek}
\usepackage{setspace}
\usepackage{graphicx}
\usepackage{amssymb}
\usepackage{subcaption}
\usepackage{tikz}
\usepackage[font=footnotesize,labelfont=bf]{caption}
\usepackage{amsmath}
\usepackage{multicol}
\usepackage{hyperref}\hypersetup{colorlinks = true, citecolor = blue, linkcolor=blue}

\usepackage{geometry}
\begin{document}
\title{\vspace{-5.5cm}\hspace*{-1.53cm}\textbf{{\mbox{{\Large Nonequilibrium steady state in a large magneto-optical trap}}}}}
\author{
\hspace*{-1.1cm}{\normalsize M. Gaudesius$^{1,2,\footnote{\hspace{0cm}Corresponding author: marius.gaudesius-1@ou.edu}}$ , Y.-C. Zhang$^3$, T. Pohl$^4$, G. Labeyrie$^1$, and  R. Kaiser$^1$} \\
\hspace*{-1.6cm}\vspace{-0.1cm}{\small\textit{ $^{1}$Universit\'{e} C\^{o}te d'Azur, CNRS, Institut de Physique de Nice, 06560 Valbonne, France }} \\ 
\hspace*{-1.3cm}\vspace{-0.1cm}{\small\textit{ $^{2}$Department of Physics and Astronomy, University of Oklahoma, Norman, Oklahoma 73019, USA }} \\
\hspace*{-1.2cm}\vspace{-0.1cm}{\small\textit{ $^{3}$MOE Key Laboratory for Nonequilibrium Synthesis and Modulation of Condensed Matter, }} \\
\hspace*{-1.2cm}\vspace{-0.1cm}{\small\textit{ School of Physics, Xi'an Jiaotong University, Xi'an 710049, China }} \\
\hspace*{-1.7cm}\vspace{-0.1cm}{\small\textit{ $^{4}$Center for Complex Quantum Systems, Department of Physics and Astronomy, }} \\
\hspace*{-1.6cm}\vspace{-0.1cm}{\small\textit{ Aarhus University, 8000 Aarhus C, Denmark }} 
 }

\date{}
\maketitle

\vspace{-0.25cm}
\leftskip=-1.15cm\rightskip=-1.15cm
{\small Considering light-mediated long-range interactions between cold atoms in a magneto-optical trap (MOT), we present numerical evidence of a nonequilibrium steady state (NESS) for sufficiently large number of atoms ($>10^8$). This state manifests itself as the appearance of an anisotropic distribution of velocity when a MOT approaches the threshold beyond which self-oscillating instabilities occur. Our three-dimensional (3D) spatiotemporal model with nonlocal spatial dependencies stemming from the interatomic interactions has recently been compared successfully to predict different instability thresholds and regimes in experiments with rubidium atoms. The behavior of the NESS is studied as a function of the main MOT parameters, including its spatiotemporal characteristics.}

\vspace*{-0.2cm}
\section{\centering\hspace{-0.38cm}{Introduction}}\label{sec:1}
\leftskip=-3.1cm\rightskip=-3.1cm

A magneto-optical trap (MOT) is known worldwide as the staple technology for producing cold and confined ensembles of neutral atoms. Conceptually simple, this device is essential in the preparation stages of highly involved tasks, e.g., Rydberg atom quantum computing \cite{extra:RydbergQC1, extra:RydbergQC2} and Bose-Einstein condensation \cite{extra:BEC1, extra:BEC2}. Additionally, the limit of a large atom number $N$ presents a wide array of interesting physics, such as random lasing \cite{0:RandomLasing1, 0:RandomLasing2}, Anderson localization \cite{0:AndersonLocalization1, 0:AndersonLocalization2}, subradiance \cite{0:Subradiance1}-\cite{0:Subradiance3}, superradiance \cite{0:Superradiance1}-\cite{0:Superradiance3} as well as spontaneous oscillations that bear similarities to plasma and stellar phenomena \cite{5:stellar1,6:MG2}. The last example falls into the category of MOT instabilities, which are, by their dynamic nature, out of thermal equilibrium. In the stable MOT regime, however, the thermodynamical considerations can be complex, requiring handling with particular care.

Stable MOTs governed by single-atom physics normally possess isotropic Gaussian velocity distributions and, thus, are deemed in thermal equilibrium. Exceptions occur in cases with Sisyphus cooling, where the distributions resemble double Gaussians \cite{extra:DoubleGaussian}. For conservative traps (such as magnetic traps or optical dipole traps, used in, e.g., Bose-Einstein condensation), the evaporative cooling processes depend on short-range collisional interactions (in particular, van der Waals interactions); after few collisions, the atoms thermalize and acquire isotropic Gaussian velocity distributions, while anisotropically trapped \cite{extra:Thermalization}. In the case studied in this paper, light-mediated long-range interactions are considered in a stable MOT containing many atoms. Compared to the case with short-range interactions, where detailed analytical results of the velocity distribution can, in principle, be obtained \cite{extra:Thermalization}, the collective forces stemming from long-range interactions (see below) pose a different challenge, making it unclear whether analytical modelling is viable. We invoke here numerical simulations and predict a velocity distribution typical of a nonequilbrium steady state (NESS), which we observe to act as a precursor to the unstable (self-oscillating) regime. In general, such a state can appear in a non-periodically driven system with broken reversibility and is characterized by the existence of nonzero net currents and a nonequilibrium probability distribution \cite{18:NESSCurrents}. States of this kind are abundant in nature, existing in processes as distant as biological ones \cite{11:Biology1,12:Biology2}.

The many-atom physics in a MOT become important typically for $N>10^4$ atoms \cite{7:Wieman}. The collective forces that appear include, e.g., the shadow force \cite{8:ShadowForce} and the rescattering force \cite{7:Wieman}. The former one is compressive and is due to an imbalance of the beam intensities in the cloud, caused by their attenuation due to the light's scattering. The latter force is, on the opposite hand, repulsive and appears as the atoms rescatter the scattered photons. These antagonistic forces are critical for understanding different features induced by multiple-scattering, e.g., the stable-cloud size increase with $N$ \cite{7:Wieman,9:SizeIncrease} as well as the occurrence of spontaneous oscillations when the beam detuning is brought close to the atomic resonance \cite{5:stellar1,10:MG1}. 

This article's main objective is to numerically study the three-dimensional (3D) particle velocity distributions of the stable regime, including how they are impacted as the unstable regime is approached. At large $N$, velocity anisotropy is observed, which is a signature of a NESS in a MOT. The NESS behavior is studied with respect to all of the main MOT parameters---the atom number $N$, the beam detuning $\delta$, the magnetic field gradient $\nabla B$ along the strong axis, and the intensity $I$ of a single beam. Moreover, given the 3D nature of the numerical simulations, we are permitted to investigate the full spatiotemporal characteristics of the NESS.

This article continues as follows. In Section \hyperref[sec:2]{II}, the theoretical model employed in the 3D simulations is briefly described as well as the details surrounding its numerical implementation. Then, Section \hyperref[sec:3]{III} presents the main results, including the identification of the NESS, its behavior versus the MOT parameters and its spatiotemporal characteristics. Finally, in Section \hyperref[sec:4]{IV}, we conclude and discuss the future perspectives.

\vspace*{-0.2cm}
\section{\centering\hspace{-0.38cm}{Theoretical approach}}\label{sec:2}

Our employed theoretical model and its numerical implementation have been detailed in Ref. \cite{13:MG3}, and here we provide a brief description. We consider the so-called \textit{balanced} MOT configuration, where the laser beams are independent and of the same, constant intensities before entering the cloud. The model is based on the hyperfine transition $F=0\rightarrow F'=1$, with each of the three Zeeman transitions $m=0\rightarrow m'=-1,0,+1$ treated as an independent 2-level system and driven by, respectively, $\sigma^-$, $\pi$, $\sigma^+$ polarized light (see Figure 1(b) in Ref. \cite{13:MG3}). The magnetic field (which splits the Zeeman levels) is linearized under the assumption of the atom position being much smaller than the radius of the MOT coils and the separation between them. The main physical effects of the model are (i) the (Doppler) trapping force; (ii) the diffusion stemming from its fluctuations; (iii) the beam intensity attenuation due to the light's scattering by the atoms; and (iv) the rescattering caused by the scattered photon exchange between the atoms. Working with a $F=0\rightarrow F'=1$ system (as opposed to a 2-level system) permits a proper description of the features related to the magnetic field and light polarization.

An important feature of our model is it being spatially nonlocal. Particularly, this is due to the atomic cross sections (in the expressions of the main effects) depending on the beam intensities, which are subject to attenuation. Note that the attenuation affects all of the main effects, including itself. In the case of the trapping force, the attenuation's inclusion results in an additional compression, i.e., the shadow force (see Figure 2(a) in Ref. \cite{13:MG3}), which is antagonistic to the rescattering (see Figure 2(b) in Ref. \cite{13:MG3}).  The attenuation naturally introduces the beam cross-saturation effect, which is computed using a devised numerical scheme. 

Moreover, the modeling is done in 3D, without any assumption of spherical symmetry. As shown and discussed more in Section \hyperref[sec:3]{III}, this allows us to observe stable clouds in different non-centrosymmetric configurations.

The complete system dynamics in our model are described by the following collisionless Vlasov-type kinetic equation for the atomic phase-space density $f(\textbf{r},\textbf{v},t)$ (refer to Eq. (26) in Ref. \cite{13:MG3}):
\begin{equation}
\label{eq:Vlasov}
\begin{aligned}
\frac{\partial}{\partial t}f+ \textbf{v}\frac{\partial}{\partial \textbf{r}}f &+ \frac{1}{M}\frac{\partial}{\partial \textbf{v}}\left\{ [ \textbf{F}_{tr}(\textbf{r},\textbf{v}) + \textbf{F}_{rs}(\textbf{r},\textbf{v}) ] f \right\} 
\\&- \frac{1}{M^2} \frac{\partial^2}{\partial \textbf{v}^2} \left[ D(\textbf{r},\textbf{v})f \right] = 0
\end{aligned}
\end{equation}
where $M$ is the atomic mass, $\textbf{F}_{tr}$ is the trapping force, $\textbf{F}_{rs}$ is the rescattering force, and $D$ is the momentum diffusion coefficient; the effects depend implicitly on time as well as on the MOT parameters mentioned earlier---$N$, $\delta$, $\nabla B$, $I$. The dependence on $N$ arises from their dependence on $I$ that contains information on the cloud density (for $\textbf{F}_{rs}$, each particle also contributes to greater repulsion), whereas the dependencies on $\delta$ and $\nabla B$ arise as these are crucial for a proper description of cooling and confining processes. Finding direct numerical solutions to this equation can be deemed impractical considering both local and nonlocal spatial dependencies. Thus, a more sensible yet an equivalent way of solving the system's dynamics is used by implementing a numerical produce based on a superparticle \cite{14:Superparticle1, 14:Superparticle2} treatment. As an important remark, note that the Vlasov equation is known to contain NESS solutions \cite{15:VlasovNESS1,16:VlasovNESS2}, meaning a MOT can potentially be found in such a state. In Section \hyperref[sec:3]{III}, we argue that a MOT indeed contains a NESS.

Our numerical implementation proceeds as follows. We start by generating a Gaussian cloud composed of $7\times10^{3}$ superparticles representing a larger amount of real particles (here, $10^{4}$ to $1.5\times10^{10}$). The cloud then evolves under the action of the effects of the model. The analysis is done by randomly selecting cloud images (here, 200, unless specified otherwise) after the initial transient time (see Figure 3 in Ref. \cite{10:MG1}). 

Note that we here concentrate on the predictions of the stable MOT regime. The unstable (self-oscillating) one has been studied with our numerical model in previous works \cite{6:MG2,10:MG1} showing qualitative agreements with the experimental observations. Nevertheless, the rationale behind picking particular MOT simulation-parameters is based on our observation, to be elaborated in Section \hyperref[sec:3]{III}, that the MOT contains a unique stable regime (NESS phase) before entering the unstable regime.

\newpage
\vspace*{-110pt}
\section{\centering\hspace{-0.38cm}{Main results}}\label{sec:3}

In this section, the main results of our simulations are presented. The 3D velocity distributions of stable MOT clouds are studied, including how they are impacted as the unstable regime is approached. The study is done with respect to all of the main MOT parameters---$N$, $\delta$, $\nabla B$, $I$---in this particular order. The spatiotemporal profiles of the 3D velocity distributions are also discussed. The analysis of our observations leads us to argue for the existence of a NESS in a MOT.

\begin{figure*}[!b]
\vspace*{-40pt}
\includegraphics[scale=0.68]{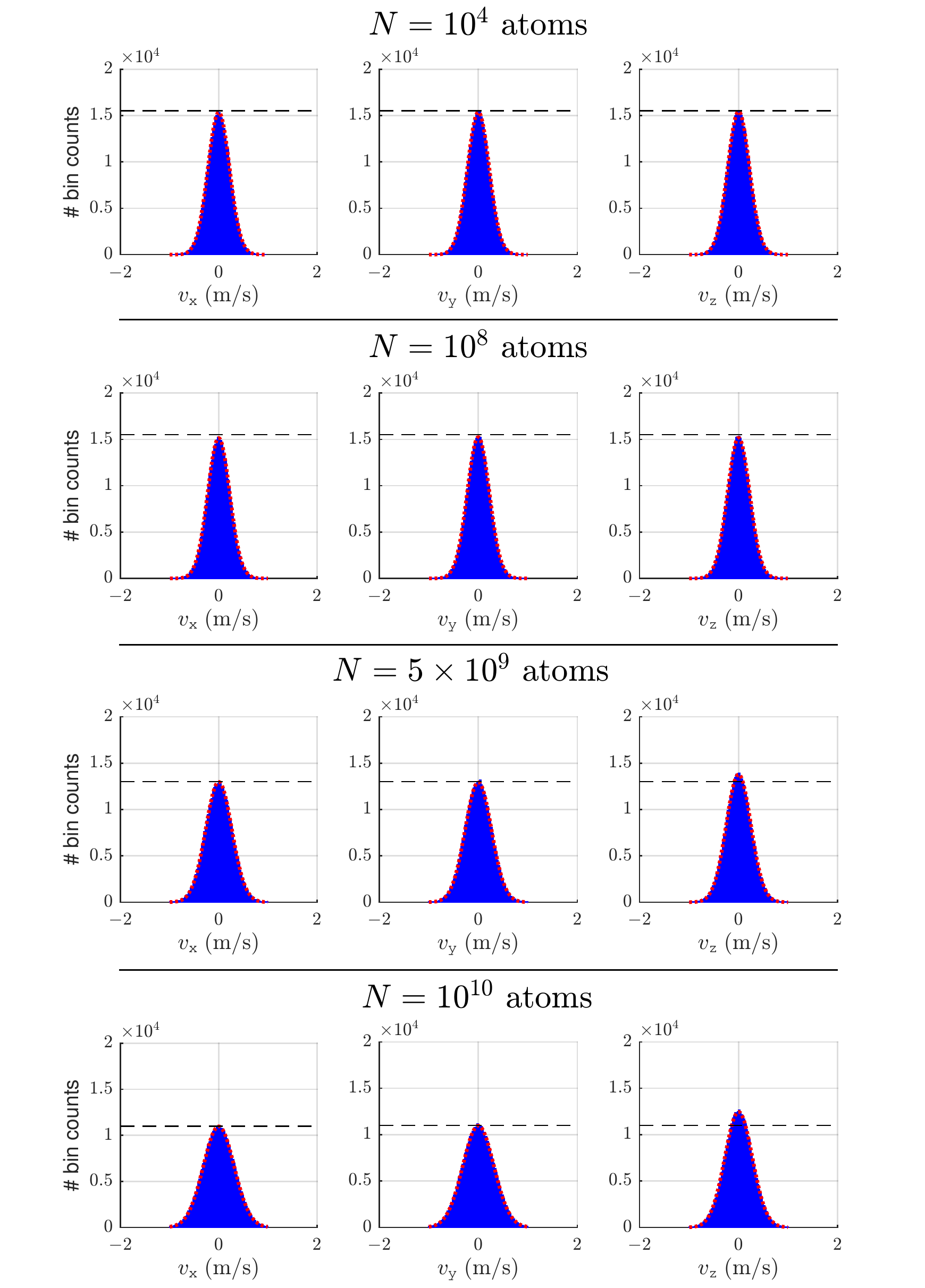}
 \captionsetup{width=1.4\linewidth}
  \caption{Histograms (blue; bin width of $25$ mm/s) for the three velocity dimensions at different values of $N$. The dotted lines are Gaussian fits. The appearance of a nonequilibrium/anisotropic velocity distribution ($v_\mathsf{z}$) above $10^8$ atoms is a signature of a nonequilibrium steady state (NESS) in a stable magneto-optical trap (MOT; dashed lines are drawn near the $v_\mathsf{x}$ distribution max). The remaining MOT parameters used in the simulations are the following: $\nabla B = 3$ G/cm, $\delta/\Gamma = -3.2$, $I=5$ mW/cm$^2$.}
\label{fig:1}
\end{figure*}

We begin by observing Figure \ref{fig:1}, showing distributions for the three velocity dimensions at different values of $N$ (the dotted lines are Gaussian fits to the histograms). The shape of the histograms is Gaussian due to the stochastic force (proportional to the diffusion coefficient) being of this kind. When above $10^8$ atoms, the distributions broaden and, more remarkably, the distribution of the $\mathsf{z}$ axis (strong magnetic field axis) narrows compared to the remaining ones. We relate these observations to the growing antagonism between the shadow force and the rescattering force (many-atom effects) as $N$ is increased. Note that different parts of the large clouds ($N>10^8$) possess varying degrees of velocity anisotropy and even have its sign switched when the atoms

\begin{figure*}[!ht]
\includegraphics[scale=0.68]{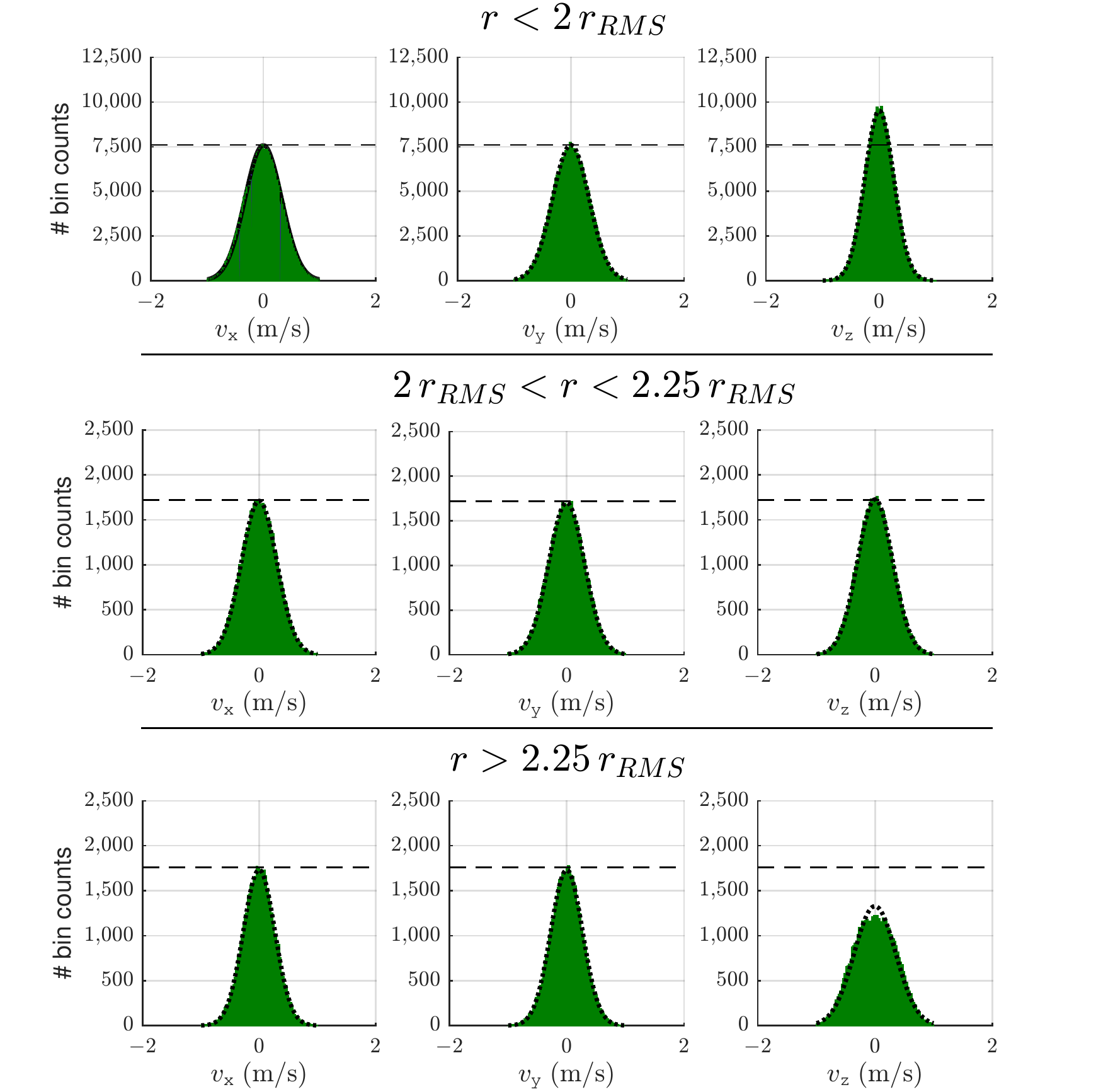}
 \captionsetup{width=1.4\linewidth}
  \caption{Histograms (green; bin width of $25$ mm/s) for the three velocity dimensions for the bottom row cloud in Figure \ref{fig:1} ($N=10^{10}$ atoms), at 3 different volumes measured from the center, with $r_{RMS}$ referring to the root-mean-square radius. 1000 cloud images were used (instead of 200). The dotted lines are Gaussian fits, and the dashed lines are drawn near the $v_\mathsf{x}$ distribution max. In the volume considered in the first row, ${\sim}70\,\%$ of all particles are concentrated, whereas for the second and third rows the concentrations are ${\sim}15\,\%$ each (hence smaller amount of bin counts). The particles in the first volume (the core particles) are responsible for the enhanced $v_\mathsf{z}$ distribution peak in Figure \ref{fig:1}, whereas the remaining ones (the edge particles) diminish it.}
\label{fig:2}
\end{figure*}

\newpage
\noindent near the cloud edge are considered. This is illustrated in Figure \ref{fig:2}, for $N=10^{10}$: The cloud core (${\sim}70\,\%$ of all particles; $r<2\,r_{RMS}$) contributes positively to the anisotropy and is responsible for the enhanced peak in Figure \ref{fig:1}, whereas the parts further away either do not contribute ($2\,r_{RMS}<r<2.25\,r_{RMS}$) or contribute negatively ($r>2.25\,r_{RMS}$) and thus diminish the overall degree of anisotropy. The change observed in the distribution widths in all the velocity dimensions (Figure \ref{fig:2}) is indicative of spatially inhomogeneous flows, as in, e.g., 3D vortexes. As discussed more when presenting the spatiotemporal velocity profiles, the appearance of a nonequilibrium/anisotropic velocity distribution is a signature of a NESS in our system and vortex-like flows can indeed form. 

\begin{figure*}[ht!]
\hspace*{30pt}\includegraphics[scale=0.7]{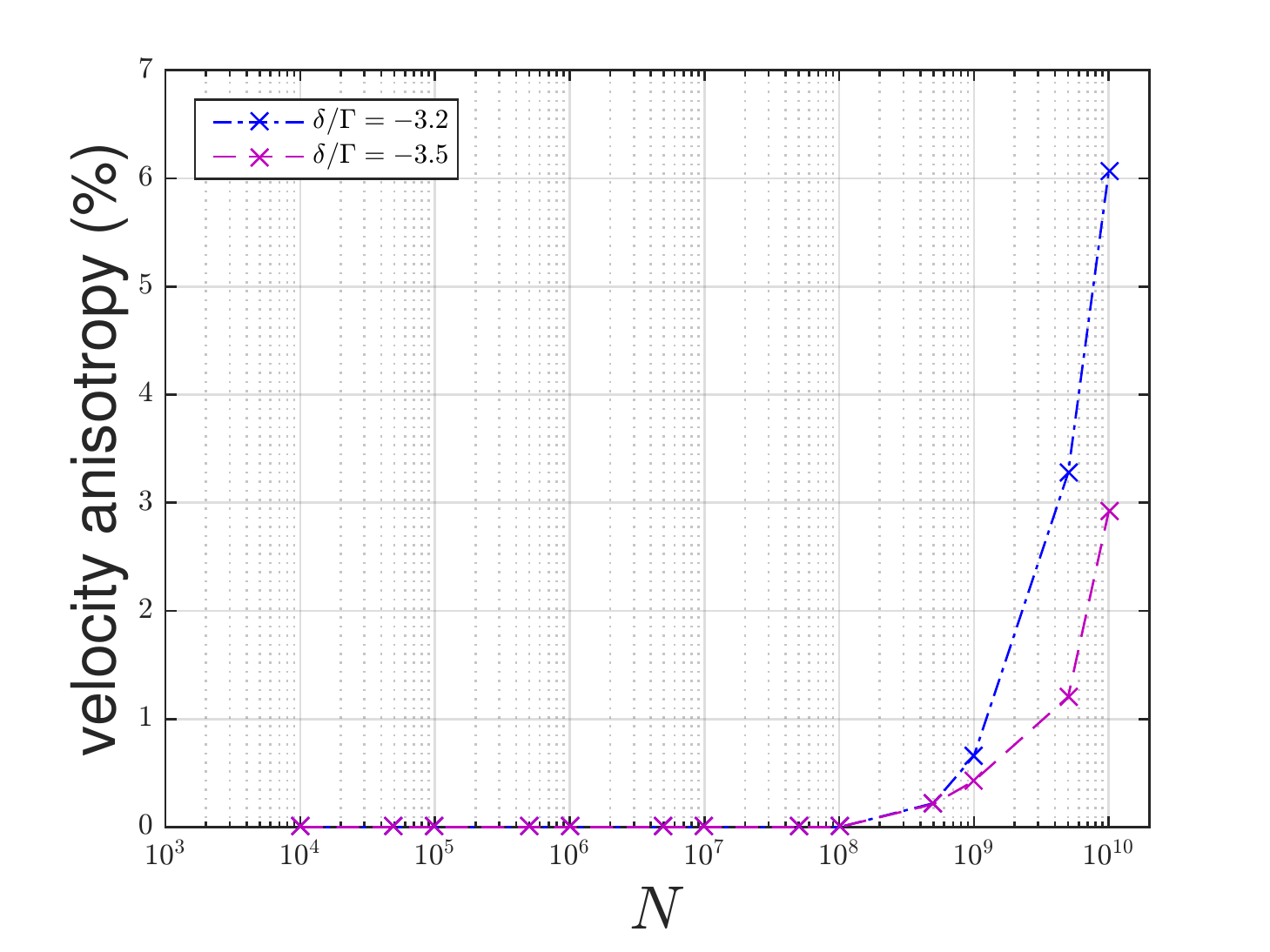}
 \captionsetup{width=1.4\linewidth}
  \caption{Velocity anisotropy versus $N$ for two different $\delta$ values. See the main text for the corresponding definition and the commentary on the displayed behaviors. Same remaining MOT simulation-parameters as in Figure \ref{fig:1}.}
\label{fig:3}
\end{figure*}

To proceed quantifying the velocity anisotropy (in whole clouds, as in Figure \ref{fig:1}), we use the normalized difference of the root-mean-square (RMS) widths of the fitted Gaussians, $\frac{\sigma_\mathsf{x}-\sigma_\mathsf{z}}{\sigma_\mathsf{x}+\sigma_\mathsf{z}}$, where $\sigma_\mathsf{x}\approx\sigma_\mathsf{y}$, $\sigma_\mathsf{z}$ are the corresponding Gaussian RMS widths. Figure \ref{fig:3} shows the velocity anisotropy plotted versus $N$ for two different $\delta$ values. Above $10^{8}$ atoms, a transition occurs from isotropic to anisotropic velocity, and the anisotropy grows as $N$ is increased but at a slower rate when $|\delta|$ is greater (at $|\delta|/\Gamma=3.5$ versus $3.2$). The transition is correlated with the unstable regime occurring above $10^{8}$ atoms, and the slower growth rate---that for a greater $|\delta|$ the instability threshold is farther \cite{10:MG1}. Below we present a natural way of extracting the NESS threshold at a \mbox{fixed $N$}.

\begin{figure*}[!h]
\hspace*{30pt}\includegraphics[scale=0.7]{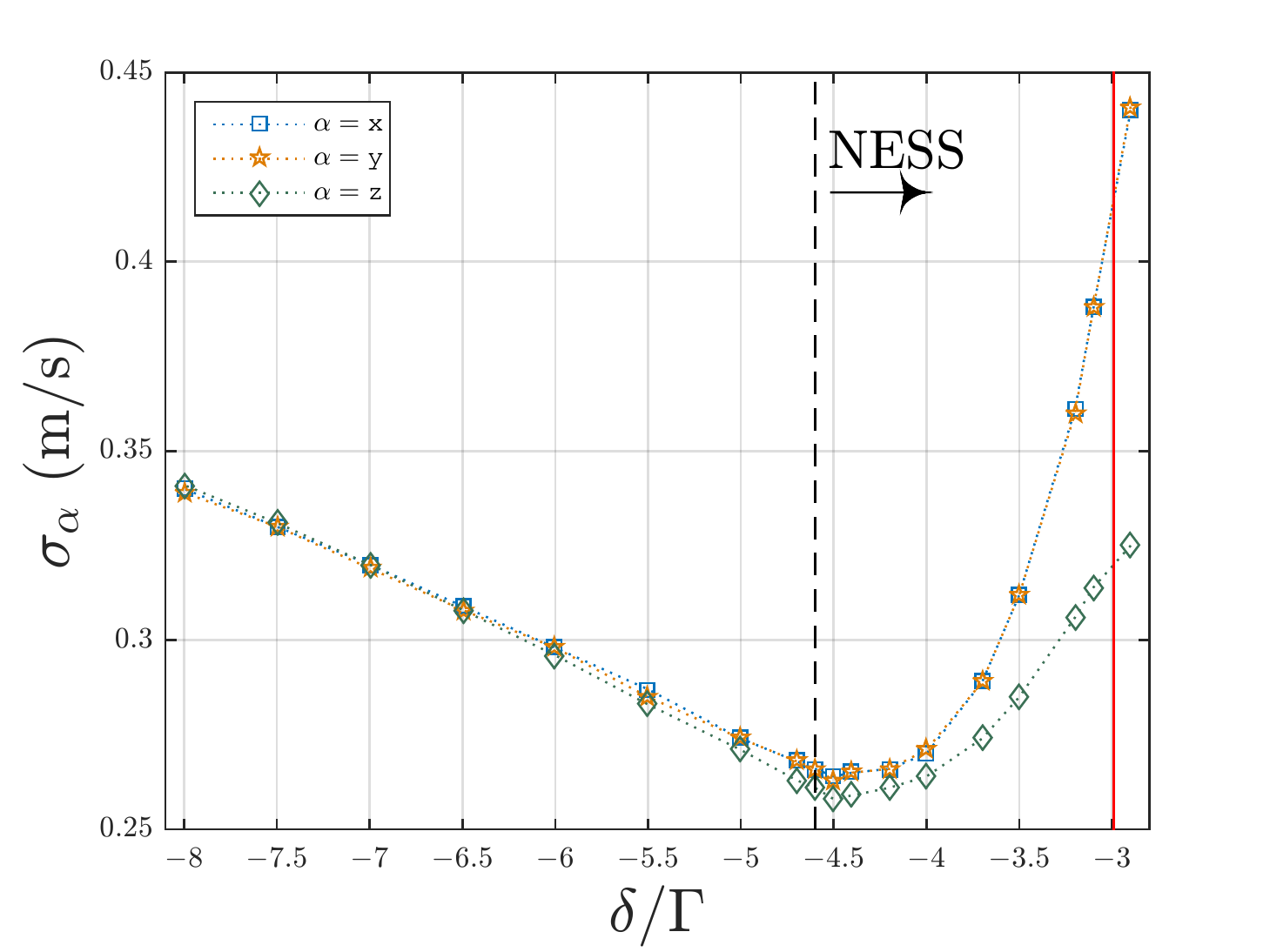}
 \captionsetup{width=1.4\linewidth}
  \caption{The root-mean-square (RMS) widths of the fitted Gaussians to velocity histograms, $\sigma_{\alpha=\mathsf{x},\mathsf{y},\mathsf{z}}$, versus $\delta$. The vertical dashed line represents the threshold beyond which the NESS behavior appears, and the vertical solid line represents the threshold beyond which the MOT becomes unstable (i.e., self-oscillations occur). The instability threshold is taken from Figure 6 in Ref. \cite{10:MG1}. Same remaining MOT simulation-parameters as in Figure \ref{fig:1}, but $N=1.5\times10^{10}$ atoms.}
\label{fig:4}
\end{figure*}

We continue by studying the NESS velocity anisotropy as the unstable regime is approached. In Figure \ref{fig:4}, we display the behavior of the RMS widths $\sigma_{\alpha=\mathsf{x},\mathsf{y},\mathsf{z}}$ versus $\delta$. As can be observed, the widths first decrease and then increase as $|\delta|$ gets lower, with the velocity remaining isotropic up to around the point, where they reach their minima. The NESS threshold occurs roughly at this point. We naturally mark it at the detuning beyond which the cloud shape diverges from ellipsoidal, to be discussed when presenting the NESS spatiotemporal characteristics. The decrease of the widths occurs due to the strengthening of the trap's friction coefficient, whereas the increase is explained by the influence of the many-atom physics. Indeed, in the single-atom limit, we find a continuous decrease of the widths up to a detuning of ${\sim}-\Gamma$, beyond which an increase is observed, in agreement with the Doppler theory \cite{extra1:DopplerLimit}. In Figure \ref{fig:4}, the anisotropy is observed to grow larger as the instability threshold is approached, which is correlated with the cloud shape changes discussed below. Note that we have added the result for the widths for the unstable cloud close to the instability threshold. We observe in this case the velocity distributions to be Gaussian and accompanying greater anisotropy than in the stable regime. We report that the distributions in the unstable regime are generally not Gaussian (close to the threshold or beyond), as different instability regimes exist \cite{6:MG2}. This is, however, beyond the scope of this paper and not discussed further.

\begin{figure*}[!h]
\hspace*{30pt}\includegraphics[scale=0.7]{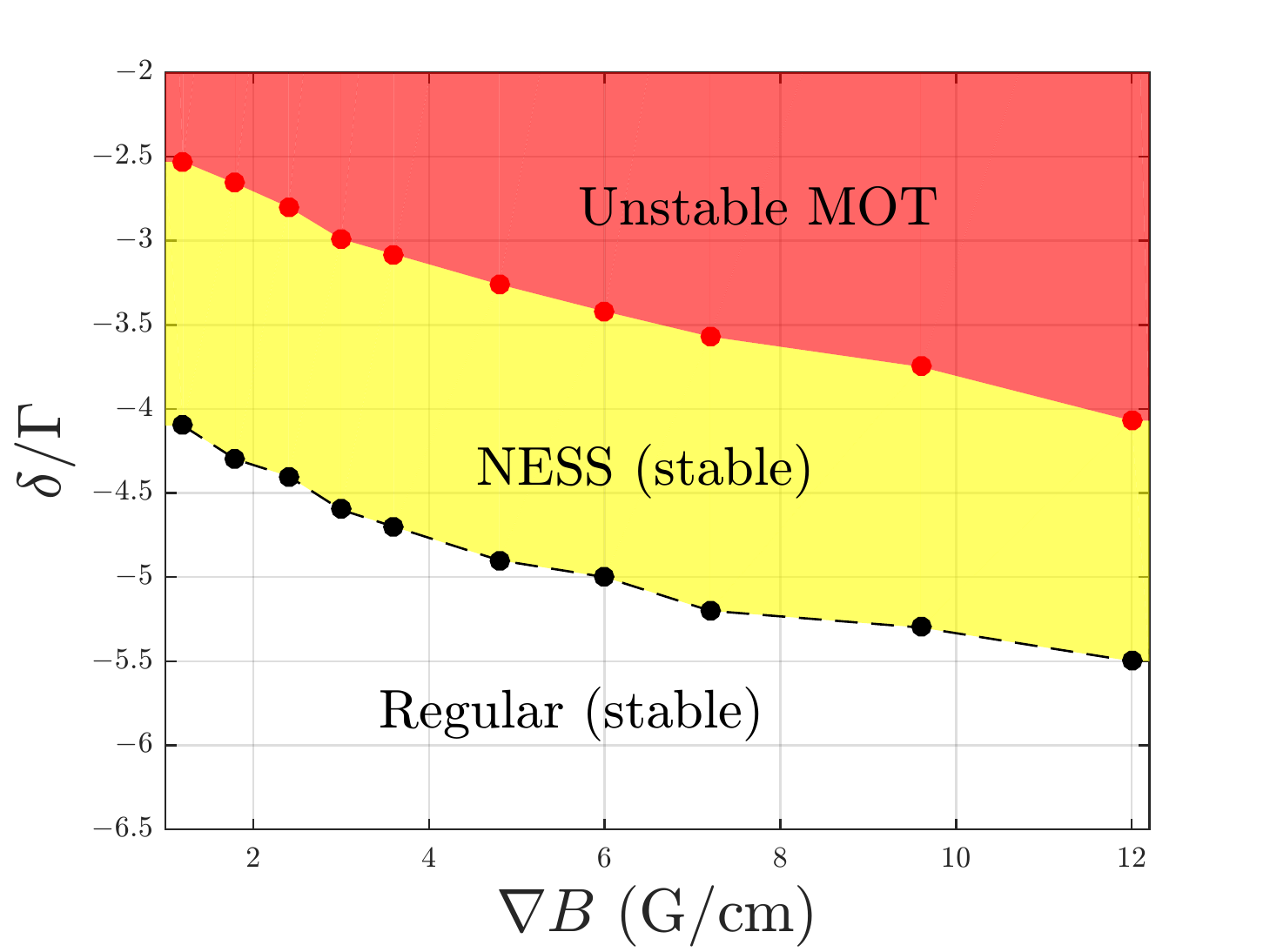}
 \captionsetup{width=1.4\linewidth}
  \caption{NESS threshold versus $\nabla B$ (dots, connected by a dashed line), superimposed with instability threshold (dots, connected by a solid line). The NESS clouds are contained between the two delineated thresholds, whereas the unstable ones are beyond the latter one. The regular stable clouds occur before the NESS clouds; unlike the latter, they possess isotropic velocity distributions and have ellipsoidal shapes (see text). The instability threshold results are taken from Figure 6 in Ref. \cite{10:MG1} (note the same data points being blue circles). Same remaining MOT simulation-parameters as in Figure \ref{fig:4}.}
\label{fig:5}
\end{figure*}

In Figure \ref{fig:5}, we plot the NESS threshold versus $\nabla B$ (dashed line), superimposed with the instability threshold (solid line; taken from Figure 6 in Ref. \cite{10:MG1}). The NESS threshold is seen to follow roughly the same qualitative behavior as the instability threshold, which decreases linearly with $\nabla B$, with the slope of $\approx -0.14\,\Gamma$ per G/cm. The negative ${\sim}1.5\,\Gamma$ offset from the instability threshold coincides with the cloud changing its shape from ellipsoidal (far-detuned case) to a different non-centrosymmetric configuration. Approaching the instability threshold closer leads to greater deformations, which we discuss when presenting the spatiotemporal profiles of the 3D velocity distributions.

\begin{figure*}[ht!]
\hspace*{30pt}\includegraphics[scale=0.7]{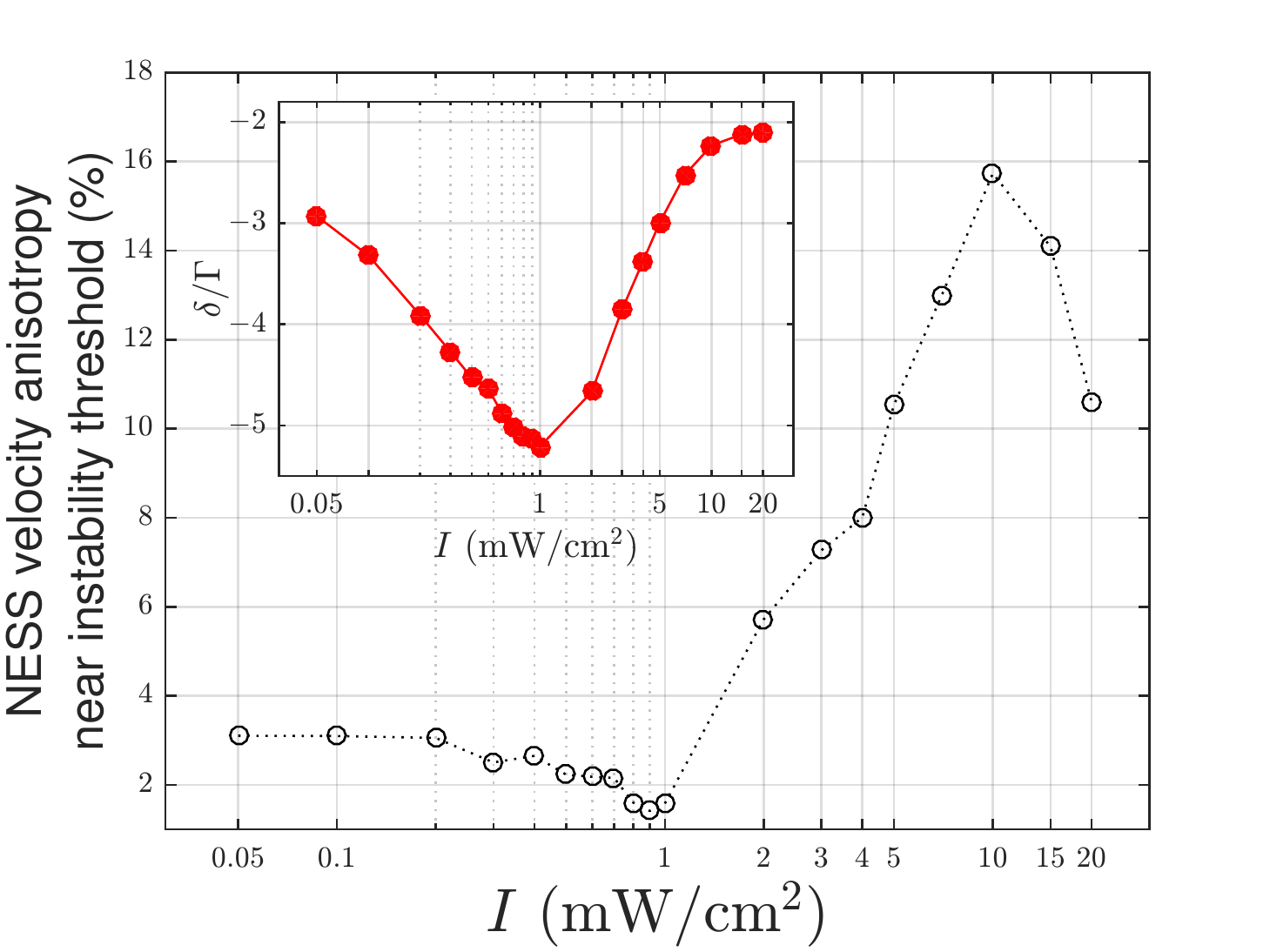}
 \captionsetup{width=1.4\linewidth}
  \caption{NESS velocity anisotropy near the instability threshold (${\sim}0.1\,\Gamma$ away) versus $I$, displayed in logarithmic scale. The circles are the data points. Same remaining MOT simulation-parameters as in Figure \ref{fig:4}. The inset displays the corresponding instability threshold results (dots) showing a resembling behavior; the results for $I\leq5$ mW/cm$^2$ are taken from Figure 3.34 in Ref. \cite{17:Thesis}.}
\label{fig:6}
\end{figure*}

We consider next examining the impact of $I$ on the NESS velocity anisotropy. In Figure \ref{fig:6}, the results versus $I$ are displayed for the NESS near the instability threshold (${\sim}0.1\,\Gamma$ away), i.e., where the anisotropy is close to being maximal (see the example Figure \ref{fig:4}). The behavior is observed to be highly non-monotonous and separated into three parts: At low $I$ ($\leq1$ mW/cm$^2$) the anisotropy proceeds decreasing, then increases at intermediate $I$ (between 1 and 10 mW/cm$^2$) and, finally, decreases again at large $I$ ($\geq10$ mW/cm$^2$). This behavior bears some resemblance to the corresponding instability threshold behavior (see the inset), which we note has been qualitatively verified by experimental results (see Figure 3.34 in Ref. \cite{17:Thesis}). This resemblance indicates that the anisotropy becomes enhanced the closer the instability threshold is to the resonance. The behaviors are, however, not exactly identical: At small $I$ the anisotropy is small (${\sim}3\,\%$) compared to the intermediate region anisotropy (growing from ${\sim}1$ to ${\sim}16\,\%$), whereas the thresholds mirror equidistantly in these regions (logarithmic scale), and at large $I$ the drop in the anisotropy (toward ${\sim}11\,\%$) corresponds to the threshold getting plateaued. Due to the NESS phase's presence being necessary for the MOT's transition into the unstable regime, this downward trend in the anisotropy size may indicate that the instability mechanism eventually gets broken for larger intensities.

\begin{figure*}[ht!]
\vspace*{-120pt}
\hspace*{0pt}\includegraphics[scale=1.03]{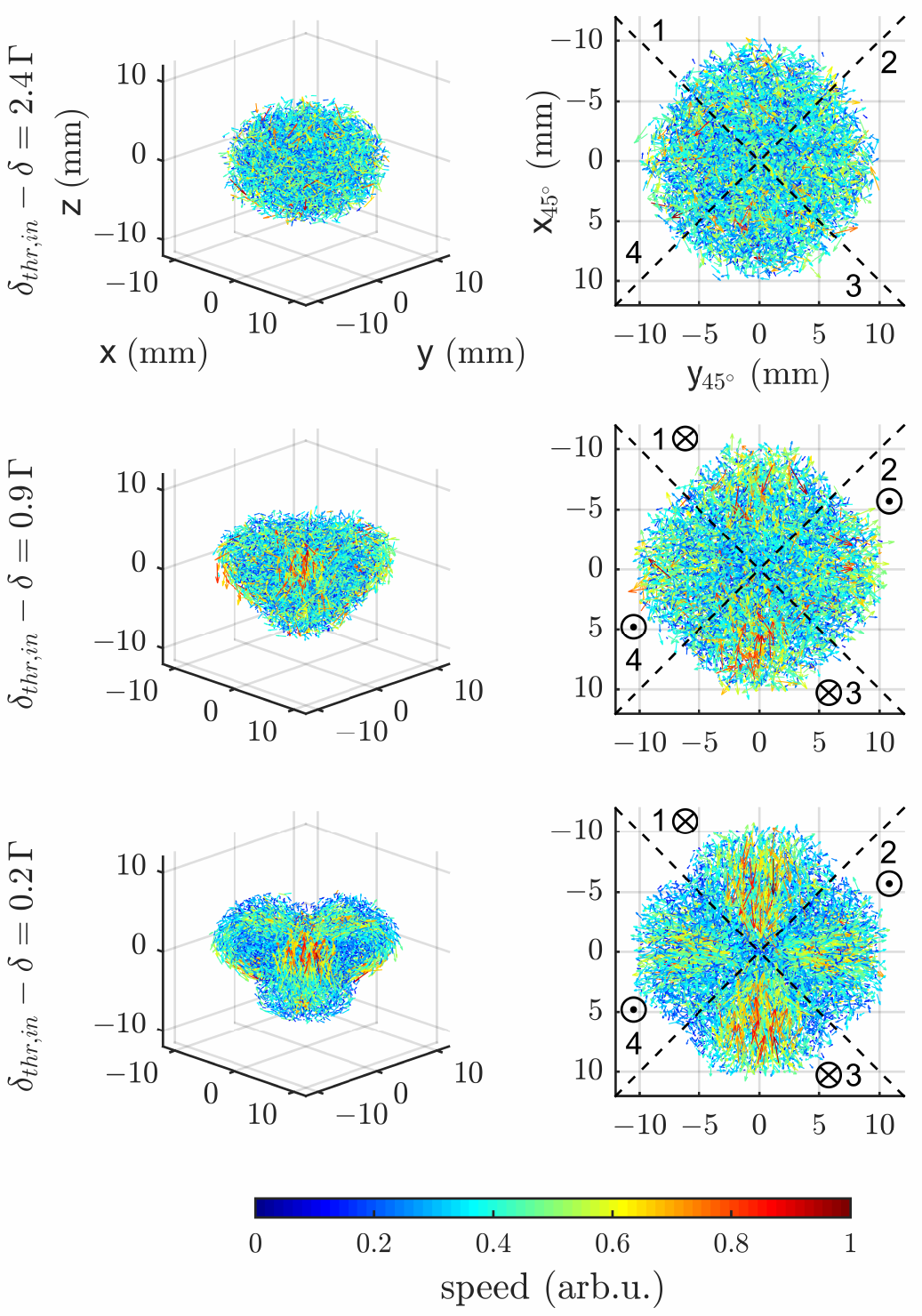}
\vspace*{-8pt}
 \captionsetup{width=1.4\linewidth}
  \caption{(Color online.) Three-dimensional (3D) velocity arrow plots for the stable clouds at different detunings from the instability threshold detuning $\delta_{thr,in}$ (rows), viewed from the side (left column) and the top (right column). Single-shot images are displayed. The $\mathsf{x}$ and $\mathsf{y}$ axes have been rotated by $45^{\circ}$ between the side- and top-view images. The diagonals in the top-view images correspond to the directions of two pairs of MOT beams (dashed lines). The clouds in the bottom two rows posses 4 lobes (separated by the beam axes); in each side-view image the lobe 1 is obscured; each top-view image shows that the lobes 1 and 3 are lower in the z plane compared to the same cloud portions in the upper-row top-view image, whereas the lobes 2 and 4 are higher. The remaining MOT parameters used in the simulations are the following: $\nabla B = 3.6$ G/cm, $I=5$ mW/cm$^2$, $N=1.5\times10^{10}$ atoms. Note that the shape changing is universal as the instability threshold is approached and accompanies growing velocity anisotropy (see the example Figure \ref{fig:4}).}
\label{fig:7}
\end{figure*}

\begin{figure*}[ht!]
\hspace*{-60pt}\includegraphics[scale=0.75]{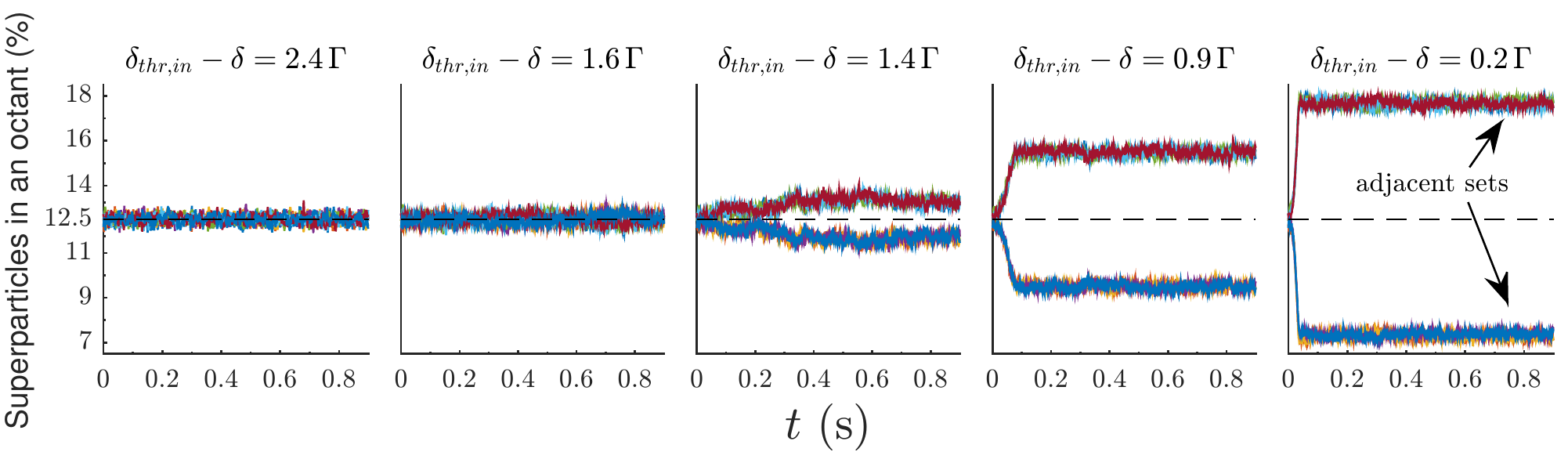}
 \captionsetup{width=1.4\linewidth}
\vspace*{-15pt}
  \caption{Evolution of the superparticle number in an octant, for the stable clouds in Figure \ref{fig:7} together with a few showing the transition to the NESS region (from $\delta_{thr,in}-\delta=1.6$ to $1.4\,\Gamma$). There are eight different-colored solid curves, each corresponding to one of the eight octants in 3D. When all the curves are on the dashed line at $12.5\,\%$, the superparticle amounts in each octant are the same. Beyond the NESS threshold, the amounts in adjacent octants become increasingly unequal. Note that the initial growths in evolutions are transients, and the information contained in such plots is used to determine the NESS threshold.}
\label{fig:8}
\end{figure*}

Finally, we discuss the spatiotemporal properties of the 3D velocity distributions of the stable clouds and visually distinguish between the \textit{regular} and NESS clouds. In \mbox{Figure \ref{fig:7}}, the velocity arrow plots are displayed for the clouds at different detunings from the instability threshold detuning $\delta_{thr,in}$ (rows), viewed from the side (left column) and the top (right column). Far from the instability threshold ($\delta_{thr,in}-\delta=2.4\,\Gamma$), the cloud takes on a flattened ellipsoid shape, as expected for a MOT having the magnetic field stronger along one of the axes (the $\mathsf{z}$ axis); the velocities are clearly disordered as the particles are subject to the effect of diffusion. Whereas this cloud is outside the NESS region (see Figure \ref{fig:5}), the other two are inside and possess noticeably different non-centrosymmetric configurations. The first one ($\delta_{thr,in}-\delta=0.9\,\Gamma$) appears triangular (side view) due to being pinched along the beam axes (top view; diagonal beam directions); the velocities start to acquire order as the particles move in a vortex-line manner around the lobes separated by the regions where the pinches occur. Closer to the instability threshold ($\delta_{thr,in}-\delta=0.2\,\Gamma$), the pinches become more pronounced and so does the order of the velocities. Figure \ref{fig:8} displays how the superparticle number in each octant is affected for the NESS region: Increasingly unequal amounts are found in adjacent octants when approaching the unstable regime, with the NESS threshold naturally set at the point before this deviation (at $\delta_{thr,in}-\delta=1.6\,\Gamma$ for the considered gradient). We observe the shape changing to be universal as the instability threshold is approached and accompanying growing velocity anisotropy (see the example Figure \ref{fig:4}). This change is indeed due to the growing antagonism between the shadow effect and the rescattering, as the unstable regime is not reached when either of the two effects are removed from the simulations \cite{13:MG3}. On an equivalent note, the stable cloud takes a flattened ellipsoid shape when attenuation (causing the shadow effect and thus fuelling the antagonism) is removed (see Figure 3.45 in Ref. \cite{17:Thesis}).  

The observation of the inhomogeneous particle-movement is important and supports the argument that a NESS is present. Indeed, as mentioned in Section \hyperref[sec:1]{I}, such a state can appear in non-periodically driven systems (e.g., our balanced MOT) and is characterized by the existence of nonzero net currents and a nonequilibrium probability distribution (which corresponds to velocity anisotropy [see Figure \ref{fig:1}]). Note that although the particles move inhomogeneously, the envelopes of the NESS region clouds nevertheless exhibit as little fluctuation as the regular stable clouds (see Figure 3.27 in Ref. \cite{17:Thesis}), proving that the definition of cloud stability is respected. 

As the last remark, there is a possibility of the NESS behavior being quasi-stationary, which otherwise has been observed in different many-body systems with long-range interactions \cite{19:QuasiStationary1,20:QuasiStationary2}. In such a case, the NESS clouds would eventually evolve into other stable configurations. The timescales for that would, however, need to exceed 5 seconds, given this is the longest duration of our simulations. For comparison, such timescales are much longer than the damping period of MOT clouds, taking some tens of milliseconds. 

\section{\centering\hspace{-0.38cm}{Conclusions}}\label{sec:4}

In this paper, we numerically predicted the existence of a NESS in a MOT containing large amount of atoms ($>10^8$). This stable-MOT state was found to emerge before the MOT's transition into the unstable (self-oscillating) regime. The NESS clouds appear in different non-centrosymmetric configurations, where the particles move in a vortex-line manner that becomes more pronounced and accompanying greater velocity anisotropy as the unstable regime is approached. This enhancement of the features results from the growth of antagonism between two collective forces---the shadow force (stemming from attenuation) and the rescattering force. Close to the instability threshold (${\sim}0.1\,\Gamma$ away), the velocity anisotropy exceeds $10\,\%$ for a wide range of beam intensities, making it possible to be detected in realistic experimental conditions. The anisotropy measurements can be further enhanced by probing the core of the clouds (refer to Figure \ref{fig:2}).  Recoil-induced resonance (RIR) spectroscopy could be employed in the velocity measurements (due to the large cloud size) \cite{extra:velocity1, extra:velocity2}, whereas the shape measurements could be achieved by cloud slicing and subsequent image reconstruction from the slices. A more systematic theoretical study of the vortex-like flows of particles might also be an interesting direction for future investigations. In this regard, conditions should arise for which the flows become turbulent, resulting in unstable cloud motion.

\section*{\centering{Acknowledgments}}

M.G., G.L. and R.K. acknowledge the financial support from the European training network ColOpt, which is funded by the European Union (EU) Horizon 2020 programme under the Marie Sklodowska-Curie action (Grant No. 721465). G.L. and R.K. further acknowledge the financial support from the European project ANDLICA, ERC advanced grant agreement (Grant No. 832219). Y.-C.Z. acknowledges the financial support from the National Key Research and Development Program Project of China (Grant No. 2021YFA1401700), the National Nature Science Foundation of China (Grant No. 12104359), and Xi'an Jiaotong University through the ``Young Top Talents Support Plan'' and the Basic Research Funding (Grant No. xtr042021012). T.P. acknowledges the financial support from the Danish National Research Foundation through the Niels Bohr Professorship and the Center of Excellence ``CCQ'' (Grant No. DNRF156).
\newline\newline
\par\hspace{171pt}\rule{4cm}{0.6pt}
\hspace{-88pt}\rule{2cm}{2pt}
\hspace{-74pt}\rule{3cm}{1.1pt}
\vspace*{-7pt}

\begin{multicols}{2}\setlength{\columnsep}{2pt}

\begingroup
\renewcommand{\section}[2]{}
\let\OLDthebibliography\thebibliography
\renewcommand\thebibliography[1]{
  \OLDthebibliography{#1}
  \setlength{\parskip}{4pt}
  \setlength{\itemsep}{1pt plus 0.3ex}
}

\endgroup

\end{multicols}


\begin{thebibliography}{9}

\small

\leftskip=-3cm\rightskip=0.15cm
\bibitem{extra:RydbergQC1}
M. Saffman, T. G. Walker, and K. Mølmer,
\textit{Quantum information with Rydberg atoms},
Rev. Mod. Phys. 82, 2313 (2010).

\bibitem{extra:RydbergQC2}
L. Isenhower, E. Urban, X. L. Zhang, A. T. Gill, T. Henage, T. A. Johnson, T. G. Walker, and M. Saffman,
\textit{Demonstration of a neutral atom controlled-NOT quantum gate},
Phys. Rev. Lett. 104, 010503 (2010).

\bibitem{extra:BEC1}
E. A. Cornell and C. E. Wieman,
\textit{Nobel Lecture: Bose-Einstein condensation in a dilute gas, the first 70 years and some recent experiments},
Rev. Mod. Phys. 74, 875 (2002).

\bibitem{extra:BEC2}
W. Ketterle,
\textit{Nobel lecture: When atoms behave as waves: Bose-Einstein condensation and the atom laser},
Rev. Mod. Phys. 74, 1131 (2002).

\hspace{20pt}\bibitem{0:RandomLasing1}
Q. Baudouin, N. Mercadier, V. Guarrera, W. Guerin, and R. Kaiser, 
\textit{A cold-atom random laser}, 
Nat. Phys. 9, 357 (2013).

\leftskip=0.15cm\rightskip=-3cm
\hspace{20pt}\bibitem{0:RandomLasing2}
L. V. Gerasimov, D. V. Kuprianov, and M. D. Havey,
\textit{Random lasing in an inhomogeneous and disordered system of cold atoms},
Optics and Spectroscopy 119, 377 (2015).

\bibitem{0:AndersonLocalization1}
F. Cottier, A. Cipris, R. Bachelard, and R. Kaiser, 
\textit{Microscopic and macroscopic signatures of 3D Anderson localization of light}, 
Phys. Rev. Lett. 123, 083401 (2019).

\bibitem{0:AndersonLocalization2}
L. A. Cobus, G. Maret, and A. Aubry,
\textit{Crossover from renormalized to conventional diffusion near the three-dimensional Anderson localization transition for light},
Phys. Rev. B 106, 014208 (2022).

\hspace{20pt}\bibitem{0:Subradiance1}
W. Guerin, M. O. Araújo, and R. Kaiser, 
\textit{Subradiance in a large cloud of cold atoms}, 
Phys. Rev. Lett. 116, 083601 (2016).

\hspace{20pt}\bibitem{0:Subradiance2}
G. Ferioli, A. Glicenstein, L. Henriet, I. Ferrier-Barbut, and A. Browaeys,
\textit{Storage and release of subradiant excitations in a dense atomic cloud},
Phys. Rev. X 11, 021031 (2021).

\leftskip=-3cm\rightskip=0.15cm
\bibitem{0:Subradiance3}
D. C. Gold, P. Huft, C. Young, A. Safari, T. G. Walker, M. Saffman, and D. D. Yavuz,
\textit{Spatial coherence of light in collective spontaneous emission},
PRX Quantum 3, 010338 (2022).

\bibitem{0:Superradiance1}
M. O. Araújo, I. Krešić, R. Kaiser, and W. Guerin, 
\textit{Superradiance in a large and dilute cloud of cold atoms in the linear-optics regime}, 
Phys. Rev. Lett. 117, 073002 (2016).

\bibitem{0:Superradiance2}
G. Ferioli, A. Glicenstein,  F. Robicheaux, R. T. Sutherland, A. Browaeys, and I. Ferrier-Barbut,
\textit{Laser-driven superradiant ensembles of two-level atoms near Dicke regime},
Phys. Rev. Lett. 127, 243602 (2021).

\bibitem{0:Superradiance3}
R. Pennetta, M. Blaha, A. Johnson, D. Lechner, P. Schneeweiss, J. Volz, and A. Rauschenbeutel,
\textit{Collective radiative dynamics of an ensemble of cold atoms coupled to an optical waveguide},
Phys. Rev. Lett. 128, 073601 (2022).

\bibitem{5:stellar1}
G. Labeyrie, F. Michaud, and R. Kaiser, 
\textit{Self-sustained oscillations in a large magneto-optical trap}, 
Phys. Rev. Lett. 96, 023003 (2006).

\bibitem{6:MG2}
M. Gaudesius, Y.-C. Zhang, T. Pohl, R. Kaiser, and G. Labeyrie,
\textit{Phase diagram of spatiotemporal instabilities in a large magneto-optical trap},
Phys. Rev. A 103, L041101 (2021).

\bibitem{extra:DoubleGaussian}
J. Jersblad, H. Ellmann, K. Støchkel, A. Kastberg, L. Sanchez-Palencia, and R. Kaiser,
\textit{Non-Gaussian velocity distributions in optical lattices},
Phys. Rev. A 69, 013410 (2004).

\bibitem{extra:Thermalization}
J. L. Bohn and D. S. Jin,
\textit{Differential scattering and rethermalization in ultracold dipolar gases},
Phys. Rev. A 89, 022702 (2014).

\bibitem{18:NESSCurrents}
D. M. Busiello, C. Jarzynski, and O. Raz,
\textit{Similarities and differences between non-equilibrium steady states and time-periodic driving in diffusive systems},
New J. Phys. 20, 093015 (2018).

\bibitem{11:Biology1}
H. Qian,
\textit{Open-system nonequilibrium steady state: Statistical thermodynamics, fluctuations, and chemical oscillations},
J. Phys. Chem. B 110, 31 (2006).

\bibitem{12:Biology2}
D. Zhang and Q. Ouyang,
\textit{Nonequilibrium thermodynamics in biochemical systems and its application},
Entropy 23, 271 (2021).

\bibitem{7:Wieman}
D. W. Sesko, T. G. Walker, and C. E. Wieman, 
\textit{Behavior of neutral atoms in a spontaneous force trap}, 
J. Opt. Soc. Am. B 8, 946 (1991).

\leftskip=0.15cm\rightskip=-3cm
\bibitem{8:ShadowForce}
J. Dalibard, 
\textit{Laser cooling of an optically thick gas: The simplest radiation pressure trap?},
Opt. Commun. 68, 203 (1988).

\bibitem{9:SizeIncrease}
A. Camara, R. Kaiser, and G. Labeyrie,
\textit{Scaling behavior of a very large magneto-optical trap},
Phys. Rev. A 90, 063404 (2014).

\bibitem{10:MG1}
M. Gaudesius, R. Kaiser, G. Labeyrie, Y.-C. Zhang, and T. Pohl, 
\textit{Instability threshold in a large balanced magneto-optical trap}, 
Phys. Rev. A 101, 053626 (2020).

\bibitem{13:MG3}
M. Gaudesius, Y.-C. Zhang, T. Pohl, R. Kaiser, and G. Labeyrie,
\textit{Three-dimensional simulations of spatiotemporal instabilities in a magneto-optical trap},
Phys. Rev. A 105, 013112 (2022).

\bibitem{14:Superparticle1}
C. K. Birsdall and A. B. Langdon, 
\textit{Plasma physics via computer simulation} (CRC, Boca Raton, FL, 2004).

\bibitem{14:Superparticle2}
T. D. Arber et al.,
\textit{Contemporary Particle-In-Cell approach to laser-plasma modelling},
Plasma Phys. Control. Fusion 57, 113001 (2015).

\bibitem{15:VlasovNESS1}
Y. Y. Yamaguchi, J. Barré, F. Bouchet, T. Dauxois, and S. Ruffo,
\textit{Stability criteria of the Vlasov equation and quasi-stationary states of the HMF model},
Physica A: Stat. Mech. App. 337, 36 (2004).

\bibitem{16:VlasovNESS2}
C. A. F. Farias, R. Pakter, and Y. Levin,
\textit{Entropy production and Vlasov equation for self-gravitating systems},
J. Phys. A: Math. Theor. 51, 494002 (2018).

\bibitem{extra1:DopplerLimit}
R. Chang, A. L. Hoendervanger, Q. Bouton, Y. Fang, T. Klafka, K. Audo, A. Aspect, C. I. Westbrook, and D. Clément,
\textit{Three-dimensional laser cooling at the Doppler limit},
Phys. Rev. A 90, 063407 (2014).

\bibitem{17:Thesis}
M. Gaudesius, \textit{Self-oscillating clouds in magneto-optical traps}, PhD thesis, Universit\'{e} C\^{o}te d'Azur, 2021, available online at \url{https://tel.archives-ouvertes.fr/tel-03273416} (accessed on 12 December 2022).


\bibitem{19:QuasiStationary1}
A. Campa, T. Dauxois, and S. Ruffo,
\textit{Statistical mechanics and dynamics of solvable models with long-range interactions},
Physics Reports 480, 57 (2009).

\bibitem{20:QuasiStationary2}
A. Gabrielli, M. Joyce, and B. Marcos,
\textit{Quasistationary states and the range of pair interactions},
Phys. Rev. Lett. 105, 210602 (2010).

\bibitem{extra:velocity1}
D. R. Meacher, D. Boiron, H. Metcalf, C. Salomon, and G. Grynberg,
\textit{Method for velocimetry of cold atoms},
Phys. Rev. A 50, R1992(R) (1994).

\bibitem{extra:velocity2}
M. Vengalattore and M. Prentiss,
\textit{Recoil-induced resonances in the high-gain regime},
Phys. Rev. A 72, 021401(R) (2005).




\end{thebibliography}
\end{document}